\documentclass[twocolumn]{aastex63}

\def\Hii    {\ion{H}{2}}

\def\apjl{\rm{ApJL}}                

\newcommand{\OII}{[\rm O~ \sc{II}]}
\newcommand{\OIII}{[\rm O~ \sc{III}]}
\newcommand{\NII}{[\rm N~ \sc{II}]}

\usepackage{graphicx}
\usepackage{lineno}                 %

\usepackage{xcolor}                 %
\usepackage{gensymb}                %
\usepackage[nointegrals]{wasysym}   %
\usepackage{amsmath}                %
\usepackage{verbatim}               %

\usepackage{color}
\usepackage{ulem} 
\usepackage{threeparttable}         %
\usepackage{bm}                     %

\begin{document}
\title{The metallicity dilution in local massive early-type galaxies}

\shortauthors{Wu \& Zhang}
\shorttitle{Dilution in local ETGs}
\correspondingauthor{Yu-zhong Wu}
\email{yzwu@nao.cas.cn}
\author[0000-0002-7670-9062]{Yu-zhong Wu}
\author{Wei Zhang}

\affiliation{CAS Key Laboratory of Optical Astronomy, National Astronomical Observatories, Beijing, 100101, China}

\begin{abstract}

We derive a sample of 114 Baldwin-Phillips-Terlevich 
diagram - star formation (BPT-SF) and Wide-field 
infrared Survey Exploer - low star formation 
rate (WISE-LSFR) early-type galaxies (ETGs) by utilizing the 
criterion W2-W3$<2.5$ 
(where W2 and W3 are the wavelengths of 4.6 and 12 
$\mu m$ in the WISE four bands) and cross-matching the $Galaxy~Zoo~1$ 
and the catalog of the Sloan Digital Sky Survey Data 
SDSS Release 7 MPA-JHU 
emission-line measurements. We find that \textbf{$\sim 28\%$} 
of our ETGs exhibit a metallicity that is at least 
2 standard deviation (0.26 dex) below the 
mass-metallicity (MZ) relation of star-forming galaxies (SFGs) 
from the SDSS. We demonstrate 
that almost all of our ETGs locate below the ``main sequence'' of 
SFGs. We find that these ETGs 
with larger metallicity deviation from the MZ relation tend to 
have lower SFR and redder color. By exploring 
the dilution properties of these massive ETGs, we report 
that the dilution effect may be mainly attributed to the inflow of 
metal-poor gas from mergers/interaction or the intergalactic medium.

\end{abstract}


\section{INTRODUCTION}

The gas-phase metallicity of the inter-stellar medium 
(ISM) of galaxies is a key parameter in galaxy formation and evolution, 
and it is regulated by injection and mixing of metals produced by star 
formation (SF), inflow of low-metalicity gas, outflow of enriched 
metallicity gas, and gas accretion from the inter-galactic medium (IGM).
A tight correlation between the stellar mass and metallicity of 
star-forming galaxies (SFGs) was first discovered by 
Lequeux et al. (1979), and this was later confirmed by observations 
of a large sample from large 
surveys (such as the Sloan Digital Sky Survey, SDSS, York et al. 2000; 
Tremonti et al. 2004, T04).

Early-type galaxies (ETGs) are often regarded as ``red and
dead'' systems at the end of galaxy evolution, where new 
stars are no longer forming. However, recent years have shown that they 
are, in fact, complicated objects. 
The presence of recent or 
ongoing star formation (SF) in some ETGs has been demonstrated by 
numberous observations (Yi et al. 2005; Kaviraj et al. 2007; 
Thilker et al. 2010; Belli et al. 2017), and this SF may originate 
from various sources.

The ultraviolet (UV) observation serves as a reliable 
tracer of recent SF in ETGs. The results from 
\textit{Galaxy Evolution Explorer} (\textit{GALEX}) found that $15\%$ of 
SDSS ETGs display 
strong UV excess (Yi et al. 2005). Using \textit{GALEX} UV imaging, 
Thilker et al. (2010) uncovered recent SF in the nearby 
ETG NGC 404, and showed that the UV light 
originates from recent SF activity 
through the analysis of archival 
images from the Hubble Space Telescope. 
The low-level SF observed in these galaxies is 
attributed to two mechanisms: external and 
internal (M\'{e}ndez-Abreu et al. 
2019; Davis et al. 2019), both of which can contribute 
significant amounts of gas to form new stars.

The external mechanism includes minor mergers with satellites
or gas-rich dwarfs (Kaviraj et al. 2009; Thilker et al. 2010; 
Ger\'{e}b et al. 2016),
major mergers with gas-rich galaxies (McIntosh et al. 2014), 
accretion of metal-poor gas from the IGM (Finkelman et al. 2011), 
and cooling gas from the outer 
halo (Lagos et al. 2014). This external 
scenario brings in gas and reignites SF. One of the most obvious events 
is a merger with gas-rich galaxies or the accretion of material from 
the IGM. In both cases, the material 
accreted is likely to have a low 
metallicity (Davis \& Young 2019).

The internal mechanism contains the stellar mass 
loss and cooling from the hot interstellar medium. 
Internal gas can be recycled to fuel low-level SF.
The material of stellar mass-loss and/or gas from the hot
phase of the interstellar medium (ISM, Sarzi et al. 2013)
can cool in massive ETGs, and then reignite
low-level SF (Pulido et al. 2018). The crucial feature of this
mechanism is that the gas, coming from the inner halo, is 
metal-rich and shares the angular momentum of the stars 
(Davis \& Young 2019).

In this work, we utilize integrated galaxy photometry from the 
catalog of the Wide-field Sky Survey Explorer (WISE, Wright et al. 2010). 
We apply a color cut of $W2-W3<2.5$ (where W2 and W3 are 
WISE bands, with central wavelengths of 4.6 and 12 $\mu m$, respectively) 
to identify WISE-LSFR (low star formation rate) ETGs. 
These ETGs are selected from galaxies located in the section of the 
Baldwin-Phillips-Terlevich 
diagram (BPT, Baldwin et al. 1981) associated 
with \Hii~ regions or star forming regions. 
Approximately $27\%$ of them exhibit a very 
low gas-metallicity relative to their stellar mass. 
Therefore,  we investigate the
properties of these ETGs. In section 2, we obtain the galaxy
sample by utilizing the Data Release 7 (DR7; Abazajian
et al. 2009) of the SDSS, and we use $\rm W2-W3<2.5$ to derive
the ETG sample from those galaxies. We present the properties of these 
ETGs, and study the main mechanism of metallicity dilution 
of our sample in Section 3. In Section 4, we
summarize our results and conclusions. Through this work, 
we assume a flat cosmology with 
$H_{\rm 0}=70~\rm km~s^{-1} Mpc^{-1}$, $\Omega_{\rm M}=0.3$, and 
$\Omega_{\Lambda}=0.7$. We adopt a Chabrier (2003) initial mass 
function (IMF).

\section{THE DATA}

The data of the SDSS DR7 serve as the 
original sample of our study.
Consequently, we employ the method of  
Wu, Zhang \& Zhao (2019) to select our
sample from the catalog of Max Planck Institute for Astrophysics --
John Hopkins University (MPA-JHU) for the SDSS DR7. 
The catalog provides comprehensive measurements of emission line fluxes, 
SFRs (Brinchmann et al. 2004), and stellar masses 
(Kauffmann et al. 2003).

We select 
galaxies within the redshift range $0.04<z<0.12$ to mitigate any bias 
arising from aperture effects (Kewley, Jansen \& Gellar 2005) and 
the redshift evolution of the MZ relation (Zahid et al. 2013). 
For all galaxies in our sample, have a covering 
fraction of more than $20\%$, which is computed based on 
the fiber and petrosian magnitudes for the r band.

For these galaxies, we apply the 
following signal-to-noise (S/N) cuts of emission lines. Since 
the $\OIII \lambda 5007$
line often appears in higher metallicity 
galaxies (Foster et al. 2012), applying 
an S/N cut to this line may 
introduce a bias in the metallicity measurements. Therefore, we 
require that the S/N ratio for H$\alpha$, H$\beta$,
$\OII \lambda \lambda 3227, 3229$, and $\NII \lambda 6584$ 
be greater than 3.
Using the BPT diagram (Baldwin et al. 1981), we 
select only those galaxies that fall within 
the H~II region on this diagram (Kauffmann et al. 2003).
Additionally, we ensure that 
the MPA-JHU catalog provides a 
measurement of the SFR for each galaxy.
For the MPA-JHU SDSS DR7 catalog, we adjust 
for the difference in initial mass functions (IMFs) 
the Kroupa (2001) IMF assumed to the 
Chabrier (2003) IMF, applying a correction factor of 
1.06. In total, 85,777 galaxies meet 
these criteria and constitute our sample.

\begin{figure}
\begin{center}
\includegraphics[width=0.6\textwidth, trim=75 0 -30 20]{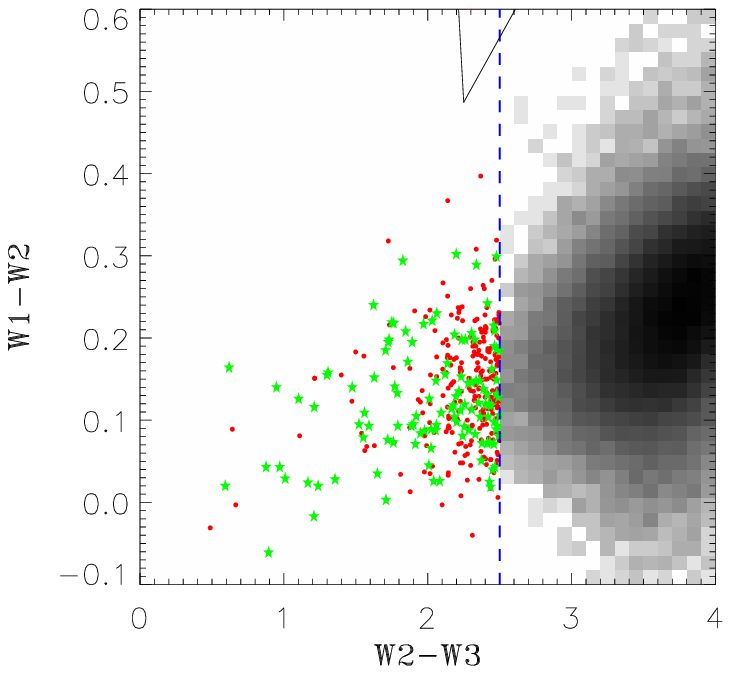}
\caption{W2-W3 versus W1-W2 Diagnostic diagram. Black dots
represent the galaxy sample with SF; red dots indicate galaxies
without SF; green pentagrams are 
confirmed ETGs. The blue dashed
line on this diagram demarcates the best boundary between galaxies 
with and without ongoing SF. The black
lines delineate the `AGN' wedge, as proposed by Mateos et al.
(2012).}
\end{center}
\end{figure}

In this work, we use the calibrator of 
T04 to estimate metallicities for our sample 
from the optical emission lines. We employ the 
$R_{23}$ metallicity indicator to calculate the oxygen 
abundances of star-forming 
galaxies (SFGs, Pilyugin, Thuan \& Vilchez 2006, 2010;
 Wu \& Zhang 2013). The SFRs are taken from the 
MPA-JHU, based on the procedure described in 
Brinchmann et al. (2004), who used resolved imaging 
to obtain a method for aperture correction 
and estimated accurate measurements of the total SFRs in galaxies.

Here, the infrared data come from the 
WISE (Wright et al. 2010), with four bands: W1, W2, W3, 
and W4, corresponding to
the central wavelengths of 3.4, 4.6, 12, and 22 $\mu m$,
respectively. 
In the WISE bands, W2 is dominated by stellar emission
(e.g. Cluver et al. 2014), while W3 includes 
strong Polycyclic
Aromatic Hydrocarbon emission bands, marking the
appearance of SF-related warm dust (Herpich et al. 2016).
Considering that the W2-W3 color is sensitive to warm dust
powered by the SF region, we use the W2-W3 color 
index of $>2.5$ and $<2.5$ to distinguish galaxies with and without 
SF activity, respectively (Herpich et al. 2016). 
We cross-match the galaxy sample with the ALLWISE source catalog 
(\dataset[doi: 10.26131/IRSA1]{https://doi.org/10.26131/IRSA1}) 
within $2''$ and require S/N$>3$ for 
W2 and W3, resulting in a sample 
of 79,028 galaxies.

\begin{figure}
\begin{center}
\includegraphics[width=0.6\textwidth, trim=75 0 -30 20]{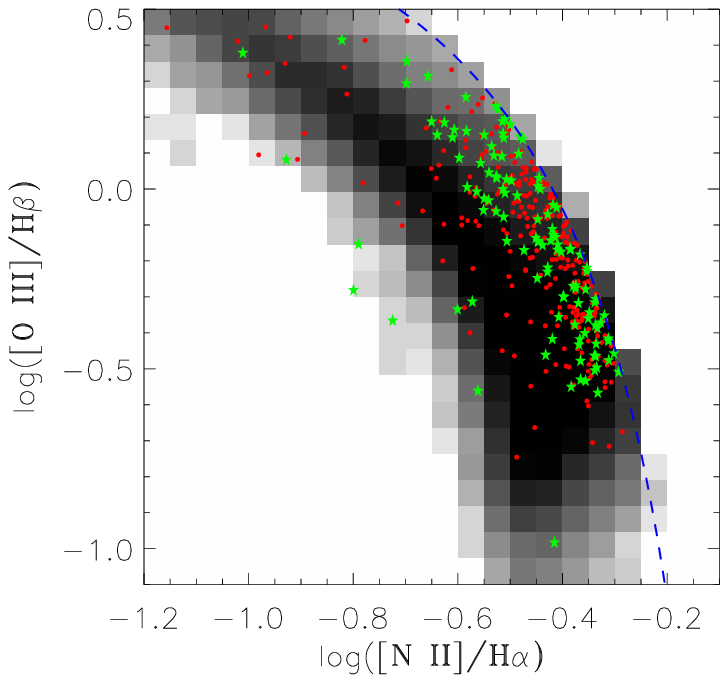}
\caption{BPT Diagnostic diagram. The blue curve shows
the Kauffmann et al. (2003) lower boundary for SFGs.
Black dots indicate the galaxy sample with SF. 
The green pentagrams are the same symbol as in Figure 1.}
\end{center}
\end{figure}

\begin{figure}
\begin{center}
\includegraphics[width=0.6\textwidth, trim=35 0 -110 20]{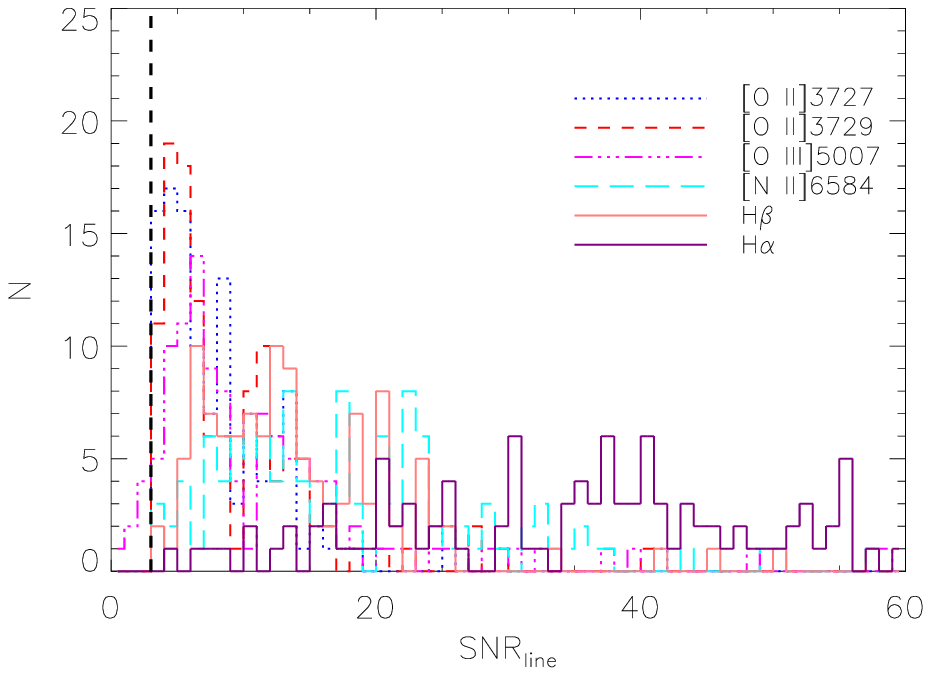}
\caption{The distributions of SNRs for various emission lines. The 
different colors and line types represent SNRs corresponding to 
different emission lines within our sample. The thick black dotted 
line shows the SNR$>3$ of various lines.}
\end{center}
\end{figure}

In this study, we first employ the 
criterion W2-W3$<$2.5 to select  
WISE-LSFR galaxies from the above-mentioned galaxies, 
and then we identify ETGs from 
these galaxies, which satisfy 
the ETG condition. For the identification 
of ETGs, we adopt two criteria: a S\'{e}rsic index
($n_{\rm sersic}$) greater than $>2.5$, and 
an elliptical probability ($p_e$) 
exceeding $>0.5$ (Herpich et al. 2018).
The measurement of the $n_{\rm Sersic}$ indexes 
are sourced from the
New York University Value-Added Galaxy Catalog 
(NYU-VAGC; Blanton et al. 2005).
The elliptical probabilities are 
derived from $Galaxy~Zoo~1$ 
(Lintott et al. 2008, 2011), where
volunteers used images of SDSS galaxies to categorize them into one 
of six types: edge-on,
clockwise spiral, anticlockwise spiral, elliptical, star/do not
know, or merger. 
Among the 79,028 galaxies, we apply the 
WISE-LSFR (defined by W2-W3$<2.5$) and ETG (based on the sersic index 
and elliptical probability) criteria to obtain 
our final sample. (1) By using W2-W3$<2.5$, we identify 391 WISE-LSFR 
galaxies from th initial sample. (2) By further applying the 
criteria $p_{\rm e}>0.5$ and $n_{\rm sersic}>2.5$, 
we ultimately obtain 117 ETGs 
that are both BPT-SF (star-forming, as indicated by their position in 
the \Hii~ region on the BPT diagram) and WISE-LSFR. 
These 117 galaxies possess three key 
characteristics: (1) they are located in the \Hii~ region on the BPT 
diagram, (2) they exhibit the WISE-LSFR property, and (3) they are classified 
as ETGs.

In this work, we have also used NUV, FUV, 
and r data to explore these galaxies. The NUV 
($\lambda_{eff}=2267$) and FUV ($\lambda_{eff}=1528$) are provided 
by \textit{GALEX}. These data originate from 
the NASA-Sloan Atlas catalog, which 
includes parameters and images 
of local galaxies derived from SDSS 
imaging and \textit{GALEX} data. 
The NSA sample ($\rm v1_{-}{0}_{-}{1}$) 
has approximatedly 640,000 galaxies. 
All the \textit{GALEX} data used in this work 
can be accessed from MAST, 
\dataset[doi: 10.17909/T9H59D]{https://doi.org/10.17909/T9H59D.}. 
In our sample, the bluer W2-W3 color indicate lower 
star formation, whereas the SDSS spectrum 
displays SF in the 
BPT diagram, and they seem to be the apparent contradiction. 
Indeed, this is not surprising, since the WISE 
color is a global measure, while the SDSS 
spectrum represents the nuclear region. 
Therefore, we present and study the 
properties of 117 BPT-SF and WISE-LSFR ETGs.

\begin{figure}
\begin{center}
\includegraphics[width=0.6\textwidth, trim=35 0 -110 20]{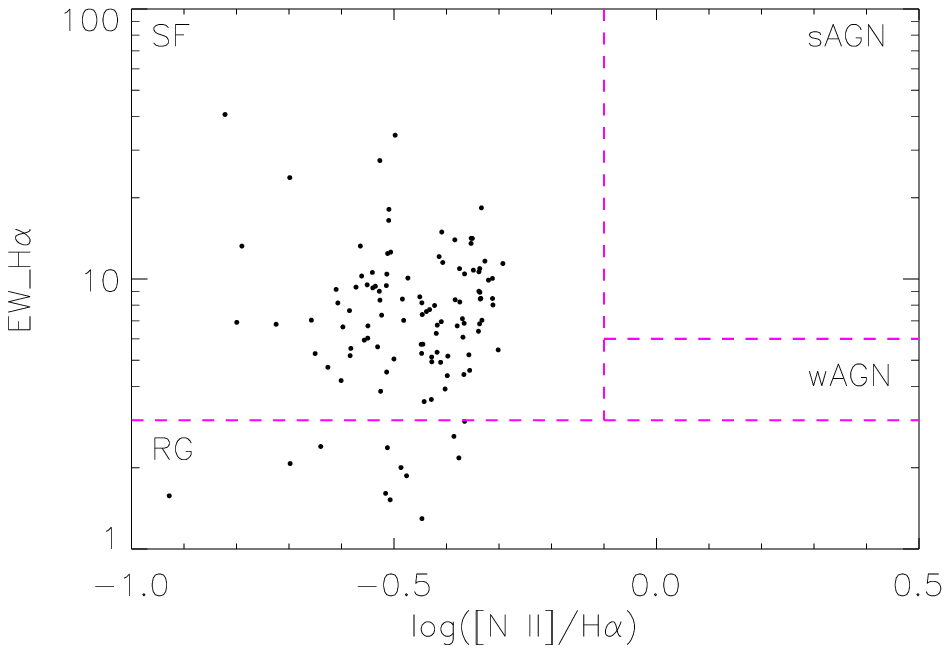}
\caption{The WHAN diagram. The magenta lines demarcate the spectral 
classes suggested by Cid Fernandes et al. 2011.}
\end{center}
\end{figure}

\begin{figure*}
\begin{center}
\includegraphics[width=0.23\textwidth, trim=400 0 -160 0]{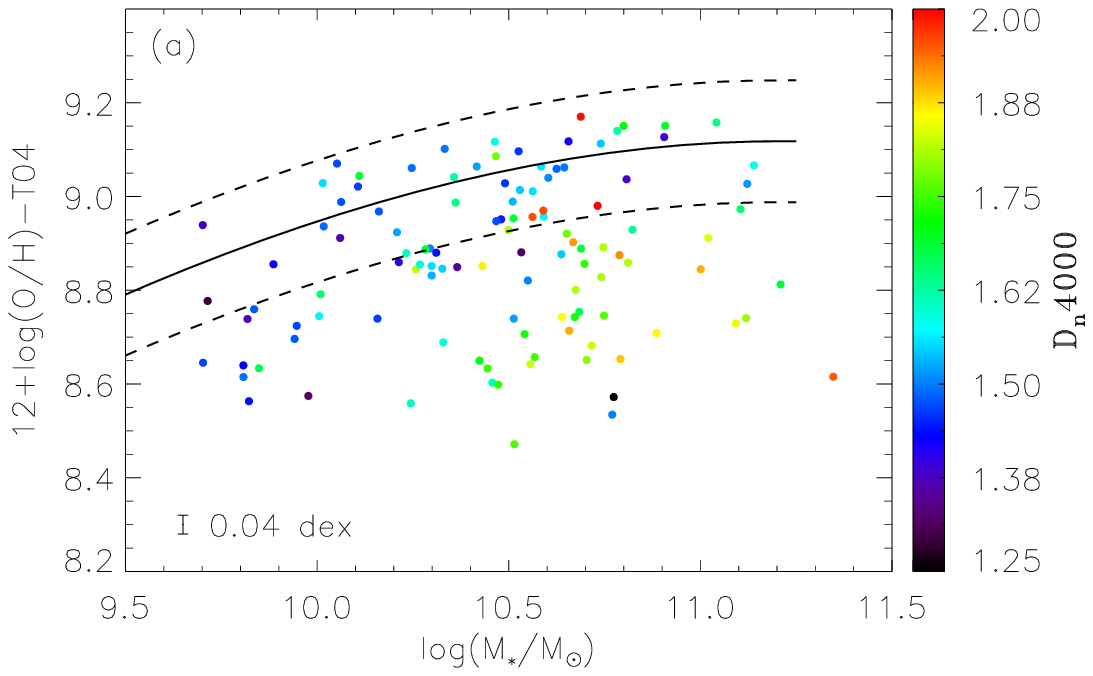}
\includegraphics[width=0.9\textwidth, trim=-530 -15 30 358]{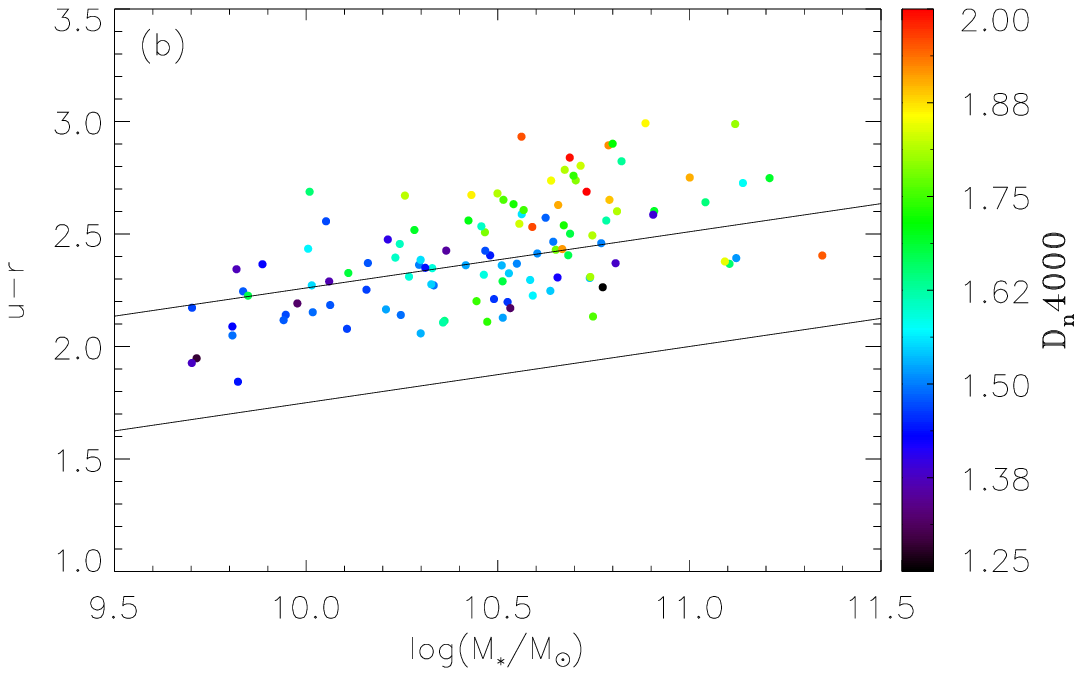}
\caption{12+log(O/H) and u-r color vs. the stellar mass with
the color bar of $D_{n}4000$. Left-hand panel: the stellar
mass vs. 12+log(O/H). The black solid and dashed curves represent the 
MZ relation and its $1 \sigma$ scatter of Tremonti et al. (2004).
Right-hand panel: the stellar mass vs. u-r color. The location of
the green valley galaxies is indicated with the two solid lines
(Schawinski et al. 2014).}
\end{center}
\end{figure*}

\section{Results}

\subsection{BPT-SF and WISE-LSFR early-type galaxies}

In Figure 1, about 99.5\% of the 
galaxies in 85777 SFG sample
are star forming, while a minority of galaxies 
without SF activity are located at the left of $\rm W2-W3=2.5$ line. 
In this Figure, 391 galaxies are BPT-SF and WISE-LSFR galaxies, 
represented by red dots, and 117 
BPT-SF and WISE-LSFR ETGs are represented 
by green pentagrams. None of these 
galaxies fall into the AGN 
region proposed by Mateos et al. (2012).
For our ETG sample, we find that the majority of
BPT-SF and WISE-LSFR ETGs possess relatively older 
stellar populations, which is associted with the 
fact that about $69\%$ of these ETGs exhibit 
$D_{n}4000 > 1.5$. Given that these ETGs reside within the 
star-forming regions of on the BPT diagram, and they are 
selected using W2-W3$<2.5$, which is a reliable
diagnostic criterion for 
distinguishing galaxies with SF and 
without SF (Herpich et al. 2016), it is 
possible that low-efficiency SF 
activity still persists 
in these ETGs. Among the 117 BPT-SF and 
WISE-LSFR ETGs, their median redshift is $\sim 0.07$.

In Figure 6 of Herpich et al. (2016), they found that
$\rm W2-W3=2.5$ is the best separator between galaxies without SF 
and with ongoing SF.
On the diagnostic diagram of W2-W3 versus W1-W2, most of our
galaxy with the WISE data lie to the right of the blue dashed
line ($\rm W2-W3=2.5$). 
In Wu (2020), they also utilized 
this color to show a separation between the ETG sample 
with SF and without SF, and they measured the metallicities of 
these ETGs with SF. In our galaxy sample, Figure 1 describes
that almost all of these galaxies have SF activities, 
and we identify a special sample 
with low-SFR through the W2-W3=2.5 color.

In Figure 2, we present the galaxy 
sample on the BPT diagram.
The blue dashed curve represents the 
lower limit for SFGs as defined by Kauffmann et al. 2003. 
The green pentagrams highlight BPT-SF ETGs
with $W2-W3<2.5$. We find that most galaxies with $W2-W3<2.5$
(shown as red dots) approach the blue line. 
Notably, a majority of BPT-SF and WISE-LSFR 
ETGs are also located close to the 
blue dashed curve
in Figure 2. This indicates that galaxy samples 
exhibiting weak SF 
activity tend to sit closer to the SFG lower 
limit of Kauffmann 
et al. (2003). Actually, 
this proximity may stem from the fact that 
SFGs in general, can receive a contribution from 
LINERs (low ionization nuclear emission-line region) given their 
typically low EW$_{\rm H\alpha}$ 
(the H$\alpha$ equivalent width).

For the spectra of our sample, we reanalyze 
the various emission lines for 117 ETG sample. 
Figure 3 shows the distributions for SNRs for the six emission lines. 
For the emission lines of H$\alpha$, H$\beta$, 
$\OII \lambda \lambda 3727, 3729$, and $\NII \lambda 6584$, 
we note that the SNRs of the five emission lines are greater 
than 3, while proportions of their SNRs being greater than 6 
are greater than $92\%$.  For $\OIII \lambda 5007$, 
we select galaxies with SNR$>2$, which does not 
have much influence on the results of this work. On the one hand, 
for ETGs, their emission lines are very weak, and the limitation 
of SNR has a significant impact on the sample size, directly 
reducing the number of samples; on the other hand, although 
the SNR reduces the sample size, it can ensure the accuracy of 
sample classification results. Therefore, the SNR of $\OIII \lambda 5007$ 
emission line is chosen to be greater than 2. 
$97\%$(110/114) of our sample have SNRs greater than 3, while 
$76\%$(84/114) of the 114 ETGS have SNR greater than 6.
From these analyses of these emission lines, we confirm 
well-detected emission lines and the BPT classification for our 
sample.

We also explore the equivalent width of H$\alpha$ 
in star-forming regions for our sample. $89\%$(101/114) of 
the 114 ETGs have EW$_{\rm H\alpha} >3$ ~\AA~, while 
$66\%$(75/114) of our sample 
have EW$_{\rm H\alpha} >6$ ~\AA~. 
Cid Fernandes et al. (2011) suggested that star-forming regions 
exhibit EW$_{\rm H\alpha} >3$~\AA~. 
In addition, we should consider our sample may be contaminated by 
LINER emission. Therefore we present 114 
ETGs on the WHAN diagram in Figure 4. According to the galaxy 
classification of Cid Fermandes et al. (2011), AGNs with 
3~\AA~$<$EW$_{\rm H\alpha}<6$~\AA~ are classified as weak 
AGNs (wAGNs), while 
AGNs with EW$_{\rm H\alpha}>6$~\AA~ are strong AGNs (sAGNs).
From Figure 4, LINERs are located in the wAGN region on the WHAN 
diagram, indicating that our sample may not be significantly 
affected by wAGN contamination. Consequently, we can ensure that 
the line ratios 
are also not contaminated by LINER emission, allowing us to obtain 
reliable metallicity measurements.

In Figure 5, we show the relations of 
stellar mass versus
12+log(O/H) and u-r color for BPT-SF and WISE-LSFR ETGs using 
the color bar of $D_{n}4000$. In Figure 5(a), the black solid 
curve is the MZ relation from Tremonti et al. (2004), and the 
black dashed lines represent its 1 $\sigma$
scatter. In Davis \& Young (2019), $7.4\%$ (42/567) of their ETGs
represents metallicities 
exceeding more than 2 standard deviation
(0.26 dex) below the Tremonti et al. (2004) MZ relation.
In contrast, $28\%$ (32/114) 
of our ETG sample deviates by at least 0.26 dex 
below the Tremonti et al. (2004) MZ relation. Furthermore, 
$56\%$ (64/114) of our ETG 
sample falls not less than 1 
standard deviation below the 
MZ relation of Tremonti et al. (2004). These findings 
indicate that our 
BPT-SF and WISE-LSFR ETG sample has a significantly 
higher percentage of galaxies exceeding 
0.26 dex below the Tremonti et al. (2004) 
relation, approximately 3.8 times 
higher than the sample of Davis \& Young (2019). 
In addition, we find that quite a few of 
these ETGs, which deviated by more 
than $1 \sigma$ standard deviation below the MZ relation of 
Tremonti et al. (2004), possess an 
older stellar population and higher stellar mass.
Compared to the metallicity difference 
at lower stellar mass,
these galaxies with higher steller masses 
show larger metallicity
differences and older stellar population.

When comparing the metallicity of ETGs and late-type   
galaxies (LTGs, also known as SFGs) at a fixed galaxy stellar mass, it 
is feasible that a statistical analysis, facilitated by the combination of 
theoretical models of galaxy formation and evolution with extensive 
data samples derived from numerous observations of these two galaxy types, 
has the potential to reveal their metallicity variations and the 
underlying causes of these differences. In Davis \& Young (2019), 
they used a large sample of ETGs to explore the origin of the gas that 
fuels RSF and considered the effect of accretion of metal-poor gas on 
ETGs. In this article, we also utilize a large sample of ETG data, 
in conjunction with theories related to galaxy evolution, to analyze 
the lower metallicity exhibited by some galaxies within our sample, and 
to delve into the possible reasons behind this phenomenon.

With regard to those ETGs with ultra-low metallicity, 
we have calibrated them using 
alternative strong line metallicity indicators, specifically 
the calibrators from Dopita (2016) and Sanders et al. (2018). 
Our findings indicate that these ETGs 
still have 
lower metallicities, confirming the reliability of 
their ultra-low metallicity status. 
When we excluded the criterion W2-W3=2.5 
and select ETGs solely based on $p_{\rm e}>0.5$ and 
$n_{\rm sersic} >2.5$, we obtain 2530 ETGs. 
Among these, $5.1\%$ (130/2530) exhibit 
metallicities that are more 
than 12+log(O/H)=0.26 dex below the MZ 
relation of Tremonti et al. (2004). Notably, 
this percentage is 
lower than that reported in Davis \& Young (2019). 
Furthermore, we observe that the percentage of ETGs 
selected with the W2-W3 cut is about 5.5 times higher 
than that without the cut. This suggests that the 
discrepancy between our 
work and Davis \& Young (2019) may stem from the fact 
that our ETGs, selected based on the W2-W3 color, 
tend to have lower SFRs compared to 
the sample used in Davis \& Young (2019).

\begin{figure}                             
\begin{center}                             
\includegraphics[width=0.6\textwidth, trim=35 0 -30 20]{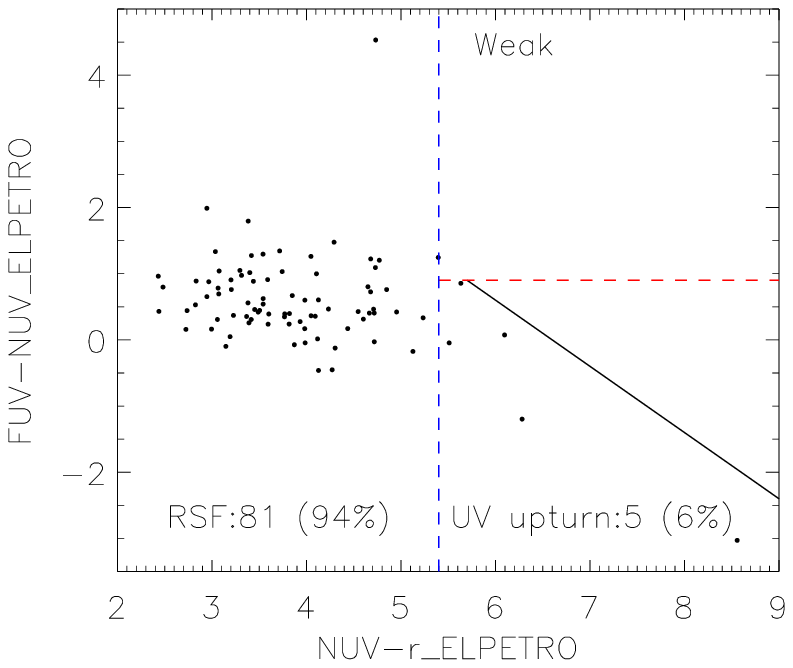}
\caption{Two-color diagram for RSF and UV upturn classification for 
our ETG sample (Yi 2011). The red horizontal dashed line represent a 
rising 
UV slope, the blue vertical dashed line denotes the criterion 
for identifying young stars, and the slanted solid line indicates the 
UV strength 
required for a galaxy to be classified as a UV upturn galaxy. RSF 
stands for residual SF.} 
\end{center}                               
\end{figure}

Figure 5(b) shows the relation of 
stellar mass and u-r color 
with the color bar of $D_{n}4000$. The two black 
solid lines represent the green valley galaxies on the stellar 
mass versus u-r color diagram, and come from equations (1) 
and (2) of Schawinski et al. (2014). From Figure 5(b), 67 ETGs 
locate in the red cloud, and more than half of 
them have an older 
stellar population ($D_{n}4000 \gtrsim 1.5$), while 47 ETGs 
lie in the green valley region. This shows that these ETGs with an older 
stellar population, higher stellar mass, and larger 
metallicity deviation from the MZ relation tend to have 
a redder color.

In Figure 6, we exhibit residual 
star formation (RSF) 
of our sample ETGs in the NUV-r 
versus FUV-NUV diagram. Due to that NUV has a high 
sensitivity for the presence of young stars, it is used to 
explore SF activity in ETGs. In fact, NUV-r $\leq5.4$ is the
most frequently used criterion to identify ETGs with RSF, and 
about 94 percent (81/86) of our 
sample galaxies met the condition.  
Based on the \textit{GALEX} and SDSS data, Jeong et al. (2022) 
suggested that the RSF of the 
sample may be from external processes (i.e. mergers or interaction), 
even if these ETGs are quiescent today.

\begin{figure}
\begin{center}
\includegraphics[width=0.6\textwidth, trim=35 0 -110 10]{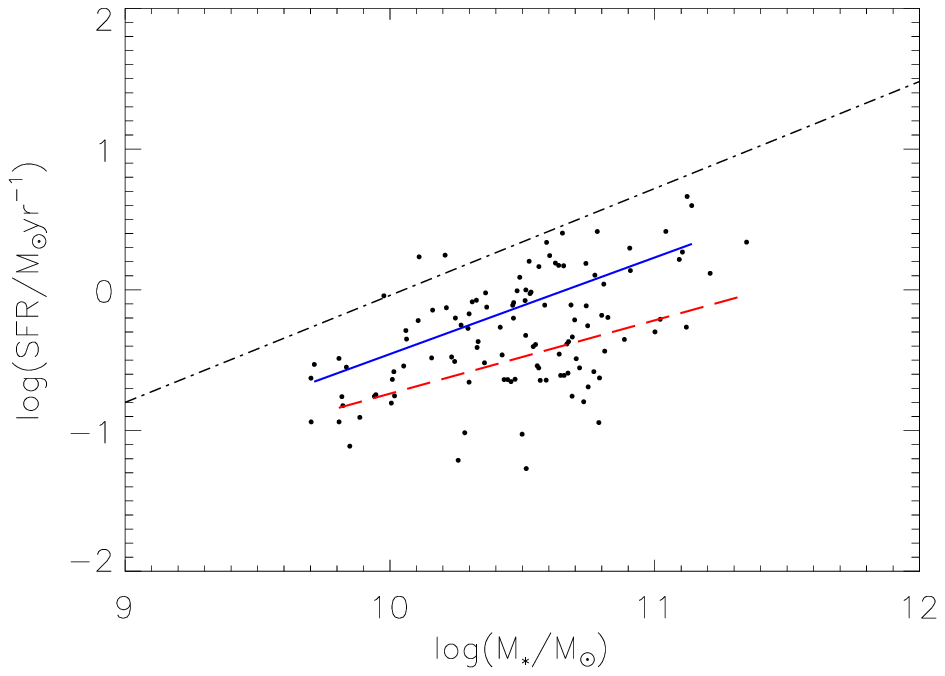}
\caption{Diagram of the stellar mass and SFR for our ETG
sample. The black dotted-dashed line represents the best fit
to the SDSS data (Renzini \& Peng 2015). The green 
pentagrams are the same symbol as in 
Figure 1. The blue solid and red
dashed lines exhibit the best least-squares fits for the two
samples, with the fits located within $1~\sigma$ region and 
below $2\sigma$ level of the MZ relation.}
\end{center}
\end{figure}

In Figure 7, we show the diagram 
of stellar mass and SFR for
BPT-SF and WISE-LSFR 
ETGs. The black dotted-dashed line is the best fit
for the SDSS data from Renzini \& Peng (2015). 114 ETGs 
are represented by the black dots in Figure 7.
 Figure 7 displays a clear correlation, indicating that galaxies with 
lower stellar masses tend to 
have lower SFRs, while those with higher stellar 
masses tend to have higher SFRs. Almost all of 
the ETGs locate 
below the fit line, except for 2 ETGs, and 
these ETGs lie below the main
sequence of SFGs by an average of 0.8 dex, showing that these ETGs
exhibit weaker SFRs. This may be 
because the W2-W3 color acts as 
a cut on the $\rm SFR-M_{*}$ diagram to remove galaxies with higher 
SFRs on the main sequence, and the remaining galaxies 
generally have lower SFRs. 
Also, we obtain the best least-squares fit 
\textbf{$\rm log(SFR/(M_{\odot}yr^{-1}))=(0.55\pm0.09)log(M_{*}/M_{\odot})-(6.08\pm0.96)$}
for our 114 ETGs. Compared with the star-forming ETG sample of 
Wu \& Zhang (2021), our sample has a significantly lower slope 
in the SFR-stellar mass relation.
In addition, the blue solid and red dashed lines represent
the best least-squares fits for the two samples, lying within the
$1\sigma$ region and below the $2\sigma$ standard deviation of
the MZ relation in Figure 5(a), 
respectively. This indicates that at a given stellar 
mass, ETGs 
with higher SFRs tend to have relatively higher 
metallicity.

\subsection{Dilution effect and low-level SF in BPT-SF 
and WISE-LSFR ETGs}

In the local study of Lagos et al. (2014), they
found that gas accretion from minor
mergers contributes neutral gas in $8\%$ of ETGs at $z=0$, while
cooling gas from the hot halos contributes the gas in $\approx90 \%$
of ETGs (depending on some model parameters). In fact, 
the most massive galaxies of ETGs
tend to host the largest hot halos (e.g., Goulding et al. 2016). 
With regard to the scheme of hot halo cooling, we examine the 
large-scale environment of our ETGs.
In Figure 8, we describe the 
number distributions of galaxies in galaxy clusters. 
The galaxy group for our ETG sample is based 
on the halo-based group finder developed by Yang et al. (2007), 
and we obtain group membership of 113 
galaxies for our ETGs. When its value is 1, it can be considered to be 
in an isolated environment. Therefore, the vast majority of 
galaxies (77/114) in our sample 
are in an isolated environment. Figure 8 shows that most of our sample 
is composed of field galaxies, indicating that 
metallicity dilution 
may not be induced by 
the hot halo cooling. In addition, Figure 8 
displays these ETGs inhabit a low-density environment (LDE), 
judged by the number of 
candidate companions, and LDE typically 
hosts ETGs that are  ``young 
and active'' compared to their cluster 
counterparts (Clements et al. 2009).

\begin{figure}
\begin{center}
\includegraphics[width=0.6\textwidth, trim=20 0 -110 5]{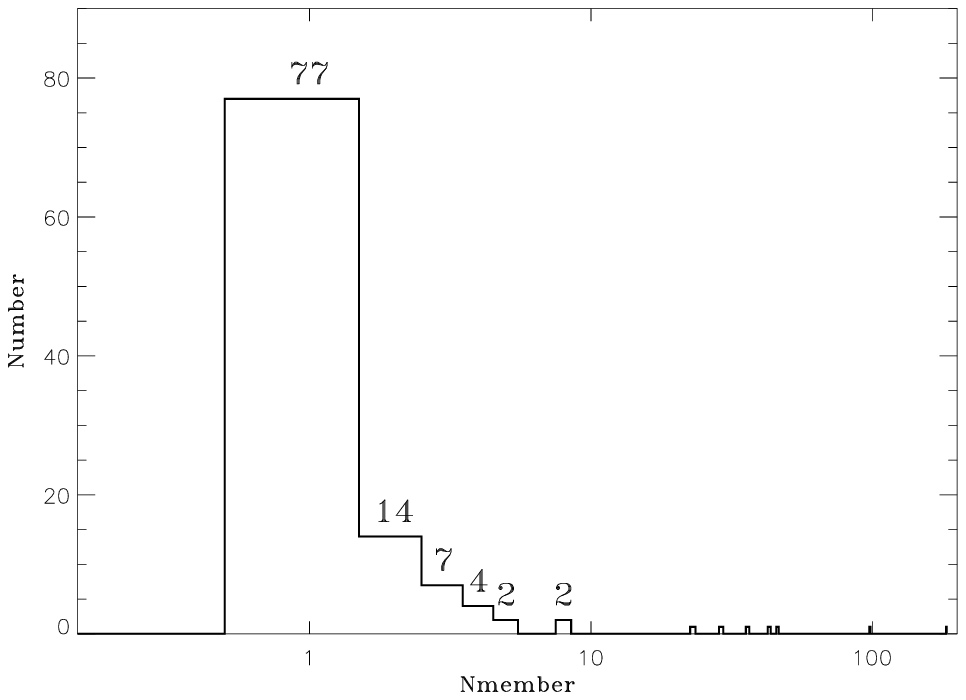}
\caption{Number distributions of galaxies in galaxy groups. 
The histograms indicate the distributions derived from 
the galaxy group sample in the SDSS.}
\end{center}
\end{figure}

Minor mergers are more common and trigger SF activity
in the local Universe (Kaviraj et al. 2009, 2014), while major
mergers are more usual at high redshift (Conselice et al. 2003).
In a sample of 42 outliers from the MZ relation of SFGs, Peeples, 
Pogge \& Stanek (2009) proposed the tidal interaction 
inducing gas inflow results in their lower metallicities. In massive 
SFG sample, Sol Alonso, Michel-Dansac \& Lambas (2010) also 
found low metallicity and high stellar mass, and suggested that 
merger-induced less enriched gas inflows produce the result.
In the two mergering galaxies of NGC4809/4810, 
Gao et al. (2023) 
found that star formation activities are triggered by the merger 
of two dwarf galaxies at the interaction area, and star-forming 
knots present poor metallicity, which may be the dilution of 
deficient-poor gas inflows during the mergering event of the 
two galaxies. Figure 5(a) shows 
the dilution effect significantly, and these 
massive ETGs have decreased $\sim 0.2-0.5$ 
dex in metallicity compared with the MZ relation of Tremonti et al.
(2004) at the same stellar mass.

The star formation in ETGs may originate from recent 
interactions or mergers, and  
indicative features of these events include shells, stellar 
streams, and tails. 
Several papers have shown that shells, tails, 
and ripples represent signatures 
of merging events (Tal et al. 
2009; Duc et al. 2015; Rampazzo et al. 2020; 
Mancillas et al. 2019). 
George (2023) supposed that star formation in blue early-type 
galaxies represents a 
stage in the evolution of ETGs. These features 
will eventually vanish, and star 
formation associated with these events 
will cease, allowing the galaxy 
to transition into another stage of 
ETGs (George 2023). In our sample, while some 
signatures of merging events 
in some ETGs may have disappeared, merger-induced star 
formation in these ETGs has indeed occurred. Therefore, 
gas inflow induced by mergers and interactions is one 
of the important reasons for 
lower metallicities in galaxies.

Because misalignment of stellar and gas
rotation in galaxies reveals the origin 
and processing of accreted gas, the fresh gas has 
more chances to appear and 
be accreted 
in ETGs, and accreted external fresh material may be 
a cause of lower SF (Davis et al. 2014; van de Voort et al. 2018). 
From observations of 1,213 galaxies of the Sydney-AAO
Multi-object Integral field spectrograph galaxy survey,
Bryant et al. (2019) also found that the misalignment fraction
is $45\pm6\%$ in ETGs. Using the MaNGA data, 
Lee et al. (2023) found that star formation 
activity in star-forming/blue ETGs comes from an external origin, 
including accretion gas from the IGM or galaxy mergers. Based on 
a finding that the stellar metallicity of ETGs is 
larger than their gas-phase metallicity, Davis \& Young (2019) 
suggested that this dilution material comes from an external
source. The redshift range of our sample is $0.04<z<0.12$,
falling within the 
sample redshift of Davis \& Young (2019).
Our sample has the lowest SFR of $0.05~\rm M_{\odot}~yr^{-1}$,
and the median value is $0.5~\rm M_{\odot}~yr^{-1}$. Our smaller 
sample likely overlaps with their 
sample, and therefore 
we suggest that the fresh gas, accreted by some ETGs in our
ETG sample, may also come from the IGM. These 
indicate that the accreted fresh gas from the IGM may be one of 
the important reasons for lower 
metallicities in galaxies.

The massive ETGs often have smaller $M_{\rm gas}/M_{*}$
ratios ($M_{\rm gas}$ is the mass of gas, Young et al. 2014), and the
inflow gas may easily dilute their 
gas-phase metallicity.
In this study, BPT-SF and 
WISE-LSFR ETGs with older stellar populations may
have a lower metallicity of 
ISM than ETGs with younger stellar populations,
and these older galaxies need less 
fresh material to dilute their
metallicities than younger ones at the same stellar mass.
We can see that BPT-SF and 
WISE-LSFR ETGs with older stellar populations
may be more easily diluted than 
those with younger stellar
populations, and most of these ETGs exceed 1 stardard deviation below
the Tremonti et al. (2004) MZ relation. Moreover, some of BPT-SF and 
WISE-LSFR ETGs with younger stellar populations also exceed 1 
standard deviation in Figure 5(a), 
and this may be
due to infalling low-metallicity gas from the external sources
having different masses, resulting in different
dilution effects and thus these 
metallicity differences.

Moreover, Figure 5 shows that galaxies with the higher
stellar mass tend to have older stellar 
populations, redder colors, and larger metallicity 
deviations from the Tremonti
et al. (2004) MZ relation.
Although we use W2-W3$<2.5$ to select our galaxies, this
color threshold primarily serves to distinguisih 
those without significant SF. 
Consequently, our sample may encompass a 
transition from galaxies with weaker SF to those 
without SF. 
These ETGs lacking SF may have experienced the inflow of 
external gas, diluting their 
gas-phase metallicity, but not start triggering 
SF. In this context, our sample 
may include galaxies with lower 
metallicity and lower SFR, while those with just-ceased SF 
activity may exhibit 
higher metallicity and higher SFR. Using 1575 MaNGA data, 
Sharma et al. (2023) found 83 galaxies 
with HI rich and low SFR, suggesting that 
a recent accretion of HI gas, without igniting SF, 
may be one of 
the key factor underlying this phenomenon.

The cold gas flow can result in the metallicity 
decreasing. Some galaxies that hold much gas may 
experience star formation, 
and their matallicity shows a relatively small 
decrease, when external fresh gas inflows. Other galaxies that 
have little gas may have 
less SF, and their metallicity exhibits a 
significant decrease, when 
external gas inflows. The two processes may correspond
to our two samples in Figure 5(a), with 
the first one falling within the
$1 \sigma$ region of the Tremonti et al. (2004) MZ relation 
in Figure 5(a), while the second one may 
lie below the $2 \sigma$
standard deviation of the MZ relation. Actually, not all inflows 
of low-metallicity material enter 
and trigger SF events, and 
the probability of this occurring is often small. 
For these older massive ETGs 
with lower gas content, a small mass of low-metallicity gas
may dilute the metallicity considerably (Belfiore et al. 2015; 
Wu 2022). In addition, we demonstrate the distribution fits
of the two samples in Figure 7, with 
the fits of the first and
second samples corresponding to the blue solid 
and red dashed lines,
respectively. Compared with ETGs with higher
metallicity, ETGs with lower metallicity are further away from
the ``main sequence'' of SFGs, and thus often have lower SFR.
This directly supports the dilution effect of metallicity 
deviation shown in Figure 5. 
Therefore, we suggest 
that these ETGs 
with metallicity dilution might be attributed to the inflow of 
metal-poor gas from mergers/interaciton 
or the IGM.

\section{Summary}

In this study, we use W2-W3=2.5 as the diagnostic tool to
derive the observational data of 114 
BPT-SF and WISE-LSFR ETGs, which cross-match 
the $Galaxy~Zoo~1$ with the galaxy sample from
the catalog of the SDSS DR7 MPA-JHU emission-line measurements.
We explore the properties of these BPT-SF and WISE-LSFR ETGs.
We summarize our results as followings:

1. We find that $\sim 28\%$ (32/114) 
of BPT-SF and WISE-LSFR ETGs exhibit a metallicity deviation of 
at least 2 standard deviation (0.26 dex) below the 
MZ relation of Tremonti et al. (2004).

2. We note that almost all of 
BPT-SF and WISE-LSFR ETGs lie below
the ``main sequence'' of SFGs proposed by Renzini \& Peng (2015)
in the SDSS data, which may be because the W2-W3 color 
is a cut on the $SFR-M_{*}$ diagram to remove galaxies with higher 
SFRs on the SFG Main-Sequence.

3. We find that a majority of BPT-SF and WISE-LSFR
ETGs sit closer to the SFG lower limit of Kauffmann et al. (2003). 
This may indicate that ETGs 
with weak SF activity may receive 
a contribution from LINERs given their generally low 
EW$_{\rm H\alpha}$. In addition, these massive 
BPT-SF and WISE-LSFR ETGs tend to have redder color.

4. As depicted in Figure 7, ETGs with a larger deviation in metallicity 
from the MZ relation of Tremonti et al. (2004) often
exhibit lower SFRs, which reinforces the 
observed metallicity dilution shown in Figure 5.

5. From the number distributions of galaxies in a group, most of 114 
ETGs are field galaxies, and this 
implies that the metallicity dilution of our ETGs 
may not be induced 
by the hot halo cooling.

6. We suggest that the dilution effect may be attributed to
the inflow of metal-poor gas from mergers/interaction or the IGM, 
and that low-level SF in most of 114 
ETGs may be attributed to the 
accretion of fresh gas in the local massive ETGs.

\section*{Acknowledgment}
We thank the anonymous referee for valuable suggestions and comments, 
which helped us to improve the paper significantly. This work is 
supported by the NSFC (No. 12090041, 12090040) and the National 
Key R/\&D Program of China grant (No. 2021YFA1600401; 2021YFA1600400).

\end{document}